\documentclass[aps,prl, floatfix,notitlepage,nofootinbib,twocolumn,superscriptaddress]{revtex4-1}
\usepackage{amsmath}
\usepackage{amssymb}
\usepackage{graphicx}
\usepackage[tight]{subfigure}
\usepackage{enumitem}
\usepackage{soul}
\usepackage{pifont}
\usepackage{xspace}

\usepackage{multirow}

\usepackage[english]{babel}
\makeatletter
\def\bbl@set@language#1{%
  \edef\languagename{%
    \ifnum\escapechar=\expandafter`\string#1\@empty
    \else\string#1\@empty\fi}%
  \@ifundefined{babel@language@alias@\languagename}{}{%
    \edef\languagename{\@nameuse{babel@language@alias@\languagename}}%
  }%
  \select@language{\languagename}%
  \expandafter\ifx\csname date\languagename\endcsname\relax\else
    \if@filesw
      \protected@write\@auxout{}{\string\select@language{\languagename}}%
      \bbl@for\bbl@tempa\BabelContentsFiles{%
        \addtocontents{\bbl@tempa}{\xstring\select@language{\languagename}}}%
      \bbl@usehooks{write}{}%
    \fi
  \fi}
\newcommand{\DeclareLanguageAlias}[2]{%
  \global\@namedef{babel@language@alias@#1}{#2}%
}
\makeatother
\DeclareLanguageAlias{en}{english}

\newcommand{\cf}{\textit{cf.} }

\newcommand{\I}{\mathbb{I}}
\newcommand{\tr}{\text{tr}}

\usepackage{bm}
\renewcommand{\vec}[1]{\boldsymbol{\mathbf{#1}}}

\newcommand{\mone}[0]{Model 1\xspace}
\newcommand{\monep}[0]{Model 1${}^{+}$\xspace}
\newcommand{\alevy}[0]{\alpha_{\text{L\'evy}} }
\newcommand{\blevy}[0]{\beta_{\text{L\'evy}} }

\usepackage[pdfusetitle,colorlinks=true,citecolor=blue,linkcolor=magenta]{hyperref}

\graphicspath{{./}{./images/}}

\begin{document}

\title{The operator L\'evy flight: light cones in chaotic long-range interacting systems}

\author{Tianci Zhou}
\email{tzhou@kitp.ucsb.edu}
\affiliation{Kavli Institute for Theoretical Physics, University of California, Santa Barbara, CA 93106, USA}
\author{Shenglong Xu}
\affiliation{Condensed Matter Theory Center and Department of Physics, University of Maryland, College Park, MD 20742, USA}
\author{Xiao Chen}
\affiliation{Kavli Institute for Theoretical Physics, University of California, Santa Barbara, CA 93106, USA}
\affiliation{Department of Physics and Center for Theory of Quantum Matter, University of Colorado, Boulder, Boulder, Colorado 80309, USA}
\author{Andrew Guo}
\affiliation{Joint Center for Quantum Information and Computer Science and Joint Quantum Institute, NIST/University of Maryland, College Park, Maryland 20742, USA}
\author{Brian Swingle}
\affiliation{Condensed Matter Theory Center, Maryland Center for Fundamental Physics, Joint Center for Quantum Information and Computer Science, and Department of Physics, University of Maryland, College Park, MD 20742, USA}

\date{\today}

\begin{abstract}

We argue that chaotic power-law interacting systems have emergent limits on information propagation, analogous to relativistic light cones, which depend on the spatial dimension $d$ and the exponent $\alpha$ governing the decay of interactions. Using the dephasing nature of quantum chaos, we map the problem to a stochastic model with a known phase diagram. A linear light cone results for $\alpha \ge d+1/2$. We also provide a L\'evy flight (long-range random walk) interpretation of the results and show consistent numerical data for 1d long-range spin models with 200 sites.


\end{abstract}

\maketitle
{\bf Introduction: }
Quantum information cannot propagate faster than light. However, in many laboratory settings, the speed of light is effectively infinite, since the natural dynamical timescales are long compared to the light-crossing time. Hence, these systems can sometimes be modeled as having instantaneous long-range interactions, for example, electric and magnetic dipolar interactions. Such non-local interactions potentially allow rapid information transfer between distant locations~\cite{guo_signaling_2019,lashkari_towards_2013,Eldredge17,Gualdi08,Avellino06}, making them attractive for quantum information processing. 

Remarkably, short range interaction enforces an emergent speed limit~\cite{Lieb1972}, even when the speed of light is effectively infinite. 
We study the analogous possibility of emergent limits on information propagation in long-range interacting systems. We refer to these limits as effective light cones even though their spacetime shape may not be that of a cone. Our focus is on power-law interactions that fall off with distance $r$ as $r^{-\alpha}$ since these systems are common in the lab and their emergent light cones have been intensely studied~\cite{Porras_2004, hastings_spectral_2006, Britton12, Blatt12,Ye13, Monroe13, gong_persistence_2014, foss-feig_nearly-linear_2015, Zaletel2015, Lukin17, Bollinger17, chen_quantum_2018,tran_locality_2018,chen_finite_2019,luitz_emergent_2019}. Using the concepts and tools recently developed from the study of many-body quantum chaos\cite{Xu_Swingle_2018,khemani_velocity-dependent_2018,chen_quantum_2018,zhou_operator_2018}, we argue that chaotic power-law interacting systems have a generic emergent light cone structure which depends only on $\alpha$ and the spatial dimension $d$.


\begin{figure}[t]
\centering
\includegraphics[width=\columnwidth]{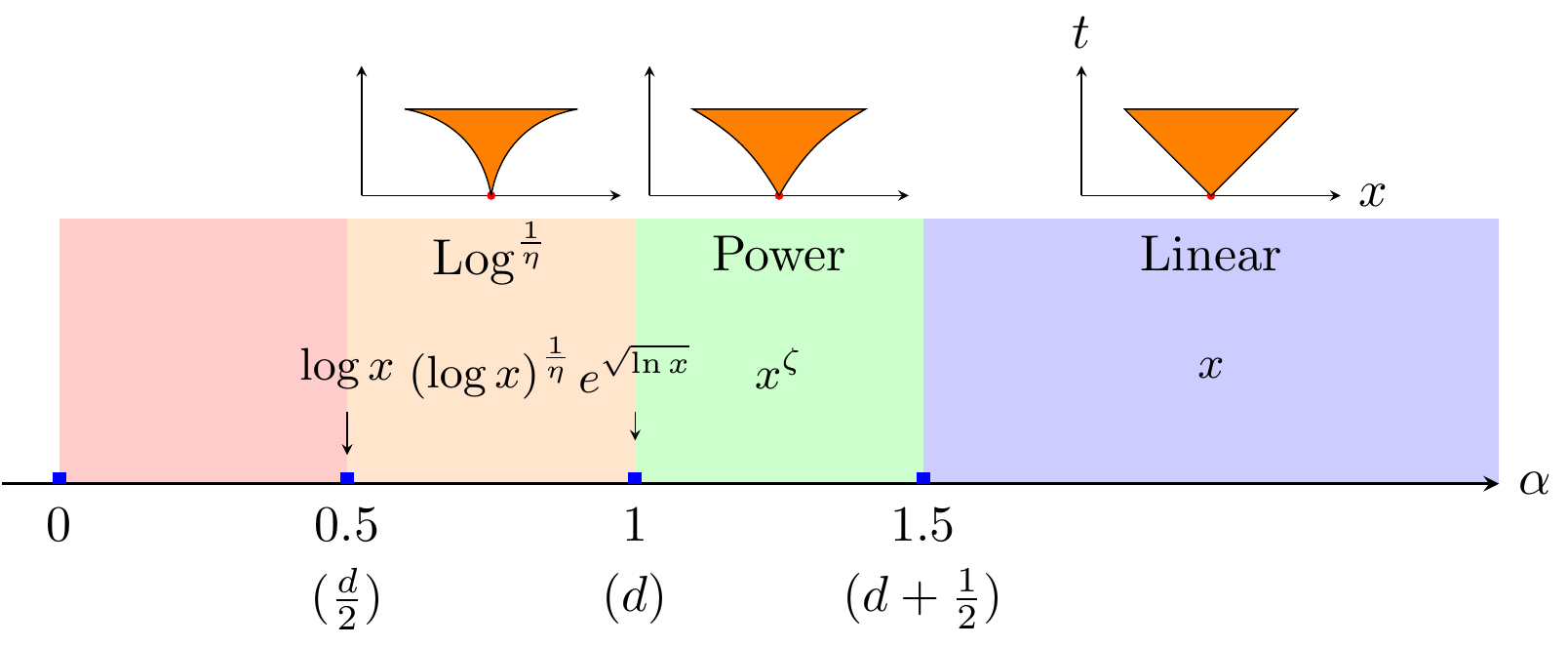}
\caption{The light cone contours of $C(x, t)$ in \mone\cite{chatterjee_multiple_2013,hallatschek_acceleration_2014}. The $\alpha$ axis marks the transition exponents in 1d ($d$-dimensional data in the parenthetical). In order of increasing $\alpha$, the light cone transitions from logarithmic to power-law to linear. The scaling functions for $t_{\rm LC}(x)$ in each phase as well as the marginal scalings at $\alpha = \frac{d}{2}$ and $d$ are displayed. The exponents $\zeta$ and $\frac{1}{\eta}$ are given by $\zeta = 2 \alpha - 2 d$, $\eta = \log_2 \frac{d}{\alpha}$. The power-law and linear light cone regimes are also numerically verified in chaotic long-range spin chains.}
\label{fig:phase_diagram}
\end{figure}

We diagnose emergent light cones by studying the commutator of two operators, where one acts as the perturbation and the other probes whether the perturbation has spread beyond a given spacetime point. 
{ Such a commutator would exactly vanish outside the light cone in a \emph{relativistic} model, whereas for quantum lattice systems without manifest Lorentz invariance, the commutator may still be nonzero for arbitrarily small times. Furthermore, for long-range interacting systems, the region outside of which the commutator is small cannot in general be bounded by a simple, \emph{linear} contour; the notion of a light cone is still applicable here, however, since information can hardly spread beyond the contour at a given point in time.}

The key quantity is the expectation value of the squared commutator (closely related to the out-of-time-ordered correlator \cite{larkin_quasiclassical_1969,kitaev2015,maldacena_bound_2015}, or OTOC) defined (in our lattice setting, at infinite temperature) as
\begin{equation}
\label{eq:C_x_t}
C(x,t)=\text{Tr}\left( [W(t),V]^\dagger [W(t),V] \right)/\text{Tr}( \I ),
\end{equation}
where $W(t) = e^{i H t} W e^{-i H t}$ is the Heisenberg form of the local operator $W$ and $V$ is another local operator a distance $x$ away from $W$. Happily, these objects can be measured in experiment~\cite{Swingle2016,Zhu2016,Halpern2016,Halpern2017,Campisi2017,Garttner2016, Wei2016, Li2017a,Landsman_2019,yao_interferometric_2016-1,yoshida_efficient_2017-1,meier_exploring_2019}, including in large-scale systems with power-law interactions~\cite{sanchez2019emergent}.

The emergent light cone is defined in terms of the spacetime contours determined by $C=\text{constant}$, as these track the effective spread of the perturbation in spacetime. For local quantum chaotic systems, one typically finds that the contours are asymptotically straight, independent of the precisely chosen contour, although in general there is a rich shape structure in the non-asymptotic regime. 
In the power-law case, Ref.~\cite{chen_quantum_2018} provided a systematic study of the light cone structure for systems with time-dependent random couplings. By random averaging, those authors gave strong numerical evidence for a complex light cone structure depending on $\alpha$.

In this work, we propose that the phase diagram in Ref.~\cite{chen_quantum_2018} is generic for chaotic power-law interacting systems {\it even without randomness}. { Specifically, we exclude systems with gauge or intrinsic constraints (see e.g. Refs.~\cite{pichler_real-time_2016,turner_weak_2018}) that prevent ergodicity.} 
Our theoretical picture is that dephasing in such systems due to quantum chaos leads to an effective stochastic description of the emergent light cone. The resulting effective model falls into the ``long-range dispersal'' class for which a universal phase diagram is known. We rigorously locate the phase boundaries that delineate the regions of ballistic, super-ballistic, and exponential growth (Fig.~\ref{fig:phase_diagram}). Furthermore, we { develop a novel numerical scheme for operator spreading using time-dependent variational principle in the matrix product representation (TDVP-MPO)\cite{Haegeman2011, Haegeman2016, Koffel2012, Hauke2013, Halimeh2017}. As far as we know, it is the most efficient method to study the operator dynamics of large scale long-range systems so far, which enables us to simulate chaotic spin chains of up to 200 sites. The results are consistent with the phase diagram in Fig.~\ref{fig:phase_diagram}.}


{\bf Operator spreading:} In general, chaotic time evolution will increase the support and complexity of $W(t)$, a process known as operator spreading.
We propose that due to dephasing, such processes can be approximated by a stochastic model that generates a universal phase diagram.

We use a height representation introduced in Ref.~\onlinecite{zhou_operator_2018,chen_quantum_2018} to describe the operator spreading, but there are many other approaches~\cite{khemani_operator_2017,nahum_operator_2018,rakovszky_diffusive_2017,von_keyserlingk_operator_2018,xu_locality_2018}. In a 1d chain of spin-$\frac{1}{2}$ particles of length $L$, we expand $W(t)$ into Pauli string basis $\{B_\mu\}$:
\begin{equation}
\label{eq:Vt}
W(t) = \sum_{\mu} a_{\mu}(t)  B_{\mu}.
\end{equation}
With the normalization $\tr( W^{\dagger}(t) W(t) ) = 1 $, the coefficients $|a_{\mu}(t)|^2$ give a normalized probability distribution over $\{B_\mu\}$.

Each basis operator has a height as follows: the $i$-th component $h_i$ for operator $B_\mu$ is $0$ if $B_{\mu}$ is identity on site $i$ and $1$ otherwise. Together these $h_i$ form an $L$-component vector $\vec h \in \{0,1\}^L$. The height representation does not distinguish different Pauli operators, so many operators have the same height. If the distribution over operators of a given height $\vec h$ is more-or-less random, then the chaotic operator dynamics is succinctly represented by the height probability distribution $f(\vec{h}, t ) = \sum_{{\rm height}(B_{\mu}) = \vec{h}} | a_{\mu}(t)|^2$. Since the commutator $[W(t),V]$ can only be non-zero if $W(t)$ is not the identity at the location of $V$, it follows that $C(x,t)$ is proportional to the {\it mean height} of $W(t)$ at site $x$ (again provided the distribution over operators of a given height is uniform).

The distribution $f$ is defined on the space of $2^L$ height states. We refer to sites with $h_i=1$ as occupied, and otherwise as unoccupied. Initially, a simple local operator $W(0)$ only has one site occupied and the distribution $f$ is concentrated on that height vector. Time evolution generally expands the operator, and the height distribution is correspondingly spread over more height configurations. Due to the decaying strength of the interaction, sites closer to $W(0)$ are more likely to increase their height earlier. As a result, the dynamics of the height distribution encodes the light cone structure.

\begin{figure}[h]
\centering
\subfigure[]{
  \label{fig:model_1}
  \includegraphics[width=0.8\columnwidth]{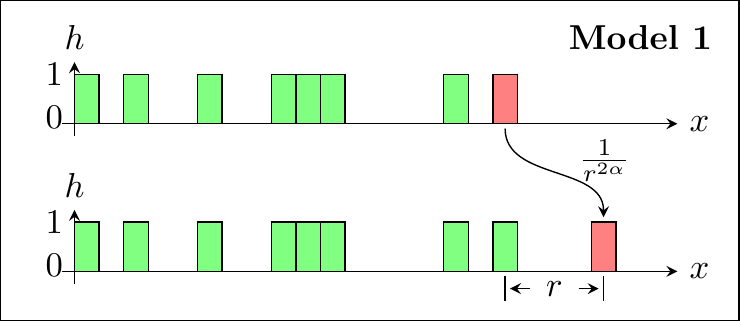}
}
\subfigure[]{
  \label{fig:model_1p}
  \includegraphics[width=0.8\columnwidth]{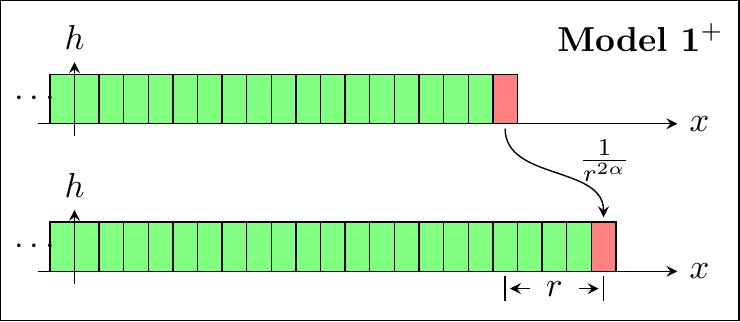}
}
\caption{\mone and a faster \monep. Filled rectangles are occupied sites. (a) Each of them (red on the top) contributes a rate proportional to ${r^{-2\alpha}}$ to occupy an empty site (red on the bottom) with distance $r$. (b) Make the same transition and then fill all the sites on its left.}
\label{fig:model_1_1p}	
\end{figure}

The height picture is particularly useful for chaotic systems because their pseudo-random character implies that the evolution of $f( \vec{h}, t )$ is often approximately Markovian. This observation has been made in local systems~\cite{khemani_velocity-dependent_2018,khemani_operator_2017,nahum_operator_2018,rakovszky_diffusive_2017,von_keyserlingk_operator_2018}, where an additional site can become occupied only if it is next to an occupied site.

We postulate the following effective Markovian transition rates for the $f$ dynamics. For definiteness, suppose the Hamiltonian is $H = \sum_\nu J_\nu H_\nu$ where the $H_\nu$ are Pauli strings with non-identity elements on only two sites a distance $r(H_\nu)$ apart and the couplings $J_\nu$ scales as $r(H_\nu)^{-\alpha}$. If the model is chaotic, then it will exhibit an effective loss of coherence on a time-scale $\tau_{\rm coh}$. The Markovian transition rates are then estimated to be of order $J_{\nu}^2 \tau_{\rm coh} \propto r^{-2\alpha}$, which leads to a probability of jumping from the top to the bottom configuration in Fig.~\ref{fig:model_1}.
Hence, the stochastic height dynamics of \mone is:
\begin{enumerate}
\item Initially only one site is occupied.
\item Each occupied site contributes a transition rate proportional to $r^{-2\alpha}$ to occupy an empty site a distance $r$ away.
\end{enumerate}

The effective dephasing and the stochastic rate estimate above are our key assumptions to understanding the light cone structure. The resulting \mone can be {\it exactly} realized in an idealized model called a Brownian circuit~\cite{zhou_operator_2018,chen_quantum_2018,xu_locality_2018}, where the couplings are Brownian motions. Here, we believe the assumed randomness of chaos can effective do the same job leading to \mone.


As discussed above, we define the light cone structure by studying its level sets of the squared commutator. The curve parameterized by $t = t_{\rm LC}(x)$ with $C(x,t_{\rm LC}(x) ) = \epsilon$ defines the light cone contour with threshold $\epsilon$, which is expected to depend strongly on $\alpha$.
In the local limit, $\alpha \rightarrow \infty$, the leading behavior is $t_{\rm LC}(x) \sim x$, i.e. a linear light cone. When $\alpha = 0$, \mone completely loses locality, and $t_{\rm LC}(x) \rightarrow 0$ in an infinite chain. The general phase diagram has been obtained exactly in Ref.~\onlinecite{chatterjee_multiple_2013,hallatschek_acceleration_2014}; translating it to our setting yields Fig.~\ref{fig:phase_diagram}.


There are four different phases characterized by different light cone scalings. In 1d, $\alpha < 0.5$ is the completely non-local phase. The transition occurs at the threshold below which the jump rate $\sim r^{-(2\times 0.5)}$ in \mone becomes un-normalizable in an infinite chain. On a finite chain, the operator spreading is similar to that of the Sachdev-Ye-Kitaev model~\cite{sachdev_gapless_1993,kitaev2015,roberts_operator_2018, zhou_operator_2018,chen_quantum_2018}. As $\alpha$ increases, one finds a phase with $t_{\rm LC}(x) \sim (\log x)^{\frac{1}{\eta}}$ ($0 < \eta \le 1$) for $0.5 \le  \alpha < 1$ and a power-law light cone phase for $1 < \alpha < 1.5$. Finally, when $\alpha \ge 1.5$, a linear light cone emerges.

{\bf A Faster Model: \monep. }
To better understand these results, and to learn more about the shape of the contours, we study an even simpler model that still captures much of the physics. We dub it ``\monep'' and illustrate in Fig.~\ref{fig:model_1p}.
Its modified transition rule is:
\begin{enumerate}[label={\arabic*$'$}]
\setcounter{enumi}{1}
\item Make a transition (as in \mone) and then fill in all the empty sites ``behind'' the newly occupied site.
\end{enumerate}

Clearly, \monep spreads faster than \mone, so its value for $C(x,t)$ will upper-bound that of \mone. However, \monep is simpler to analyze because its state is completely determined by the motion of the outer-most point, thus reducing it to a single particle problem. In 1d, the dynamics can be sped up by taking all the sites with $x \le 0$ to be occupied in the initial height state. The motion of the outer-most point becomes Markovian, and the rate to move forward $r$ sites is then $\sum_{r'=-\infty}^r (r')^{-2\alpha}  \sim r^{1-2\alpha}$.

Such a long-range random walk is called a L\'evy flight (see Refs.~\onlinecite{calvo_generalized_2010,janson_stable_2011,chechkin_introduction_2008}), where the displacement of each jump $X_t$ (at time $t$) is an independent random variable with distribution $f_{\rm jump}(x)$ that scales as ${x^{-(1 + \alevy)}}$ when $x \rightarrow \infty$. According to the generalized central limit theorem \cite{SM},
the total displacement will converge to a L\'evy stable distribution $L_{ \alevy, \blevy}$, with parameter $\alevy = 2\alpha - 2$ and $\blevy = 1$ for the present case.
The distribution for the right-most occupied site $\rho( r, t )$ scales as
\begin{equation}
\label{eq:rho_monep}
\rho( x, t )
\sim
\left\lbrace
\begin{aligned}
  & L_{2\alpha - 2, 1} \left({x}/ {t^{\frac{1}{\zeta}}}\right) & \,\,  1 < \alpha \le  1.5, \\
  & L_{2\alpha - 2, 1} \left({(x - v_B t)}/{t^{\frac{1}{\zeta}}}\right) & \,\,  1.5 < \alpha < 2, \\
  & \exp\left( - {(x - v_B t)^2}/{2Dt} \right) & \,\, 2 \le \alpha,
\end{aligned} \right.
\end{equation}
where $L_{\alpha, \beta}$ is the L\'evy stable distribution\label{app:GCLT} $\zeta = 2\alpha - 2$ and $v_B $ and $D$ are the first and second moments of $f_{\rm jump}(x)$ when they exist. The probability for site $x$ to be occupied is equal to $\int_x^\infty \rho(x', t )\,dx'$ in \monep, which leads to the light cones in the second column of Table ~\ref{tab:op_lightcone}:

\begin{table}[h]
\centering
\begin{tabular}{ |c|c|c|c|c|c|c| }
 \hline
  & \multicolumn{3}{c|}{\text{\monep}}  & \multicolumn{3}{c|}{\text{\mone}} \\ \hline
  $\alpha$ & LC & width & tail  & LC & width & tail \\ \hline
  $ [0.5,1) $ & N/A &  N/A & N/A  & $e^{t^{\log_2 \frac{1}{\alpha}}}$ & N/A & \multirow{2}{*}{\parbox{35pt}{\vspace{0cm} $x^{-2\alpha}$ }} \\ \cline{1-6}
  $ (1,\frac{3}{2}] $ & $t^{\frac{1}{2\alpha - 2}}$ & N/A & \multirow{2}{*}{\parbox{35pt}{\vspace{0cm} $x^{-(2\alpha - 2)}$ }} & $t^{\frac{1}{2\alpha - 2}}$ & N/A & \\ \cline{1-3} \cline{5-7} 
  $ (\frac{3}{2}, 2) $ & \multirow{3}{*}{\parbox{15pt}{\vspace{0cm} $v_B t $ }} & $t^{\frac{1}{2\alpha - 2}}$ & &\multirow{3}{*}{\parbox{15pt}{\vspace{0cm} $v_B t $ }} & $t^{\frac{1}{2\alpha - 2}}$ &  $x^{-(2\alpha - 2)}$  \\ \cline{1-1} \cline{3-4} \cline{6-7}
  $ 2 $ &  & $( t\ln t )^{\frac{1}{2}}$ & \multirow{2}{*}{\parbox{35pt}{\vspace{0cm} Gaussian }} & & $( t\ln t )^{\frac{1}{2}}$ & \multirow{2}{*}{\parbox{35pt}{\vspace{0cm} Gaussian }} \\ \cline{1-1} \cline{3-3} \cline{6-6}
  $  (2,\infty) $ & & $t^{\frac{1}{2}}$ &  & & $t^{\frac{1}{2}}$ & \\ \hline
\end{tabular}
\caption{Scalings of light cone, its broadening (width) and tail of \monep and comparison with \mone.}
\label{tab:op_lightcone}
\end{table}

{ The transition points $\alpha = 1, 1.5$ and $2$ are the critical values above which the jump distribution $f_{\rm jump}(x)$ of \monep starts to be normalizable and acquires mean velocity $v_B$ and variance $D$ respectively. In the following, we review the quantitative predictions on \mone by \monep. Aside from the light cone scalings and characteristic width, we also study the wavefronts' spatial dependences at fixed time. We refer to the large-$x$ limit of $C(x,t)$ at fixed $t$ as the \emph{tail}.  For small $t$ in \mone, the tail should be roughly equal to the probability of a rare jump from the initial seed at site $0$, i.e. as ${x^{- 2\alpha } }$. The tails we discuss are for large $t$.

From Tab.~\ref{tab:op_lightcone}, all the scalings about the light cones are identical for both models when $\alpha \ge 1.5$. In this regime, \monep has a linear light cone and since it spreads faster than \mone, the later must also have a linear light cone. We would further expect \mone to form a domain of occupied sites within the light cone, rendering the two models qualitatively similar. In particular the widths of ${t^{1/({2\alpha - 2})}}$ and $\sqrt{t}$ have been verified in the classical simulation of \mone \cite{SM}.

When $1 < \alpha < 1.5$, \monep has a power-law light cone, whereas that of \mone could potentially be more restrictive. But suppose \mone were to have a linear light cone; then a domain of occupied sites would form, so that the light cone of \mone would be identical to that of \monep. But the latter has faster-than-linear propagation, leading to a contradiction. In practice, \mone has the same light cone scaling as \monep \cite{chatterjee_multiple_2013,hallatschek_acceleration_2014}, but the gaps between filled sites in \mone gives a different tail scaling than \monep. Within a mean-field approximation \cite{SM}, we find the tail scaling to be ${x^{-2\alpha}}$, which is further numerically verified in \mone and a long-range spin chain discussed below.

Finally, when $\alpha < 1$, the the long range jumps of \mone create large gaps between the occupied sites. The approximation of a solid domain as in \monep does not work, and the problem is many-body in nature.}

We briefly comment on the situation in higher dimensions. The transition rate ${r^{-\alpha}}$ is normalizable in $d$-dimension only when $\alpha > \frac{d}{2}$. When we consider the corresponding \monep, the outer-most point jumps with rate $\int d^dr\; {r^{-2\alpha}} \sim r^{-2\alpha +d} $. The existence of the zeroth, first and second moments gives the general transition points marked in Fig.~\ref{fig:phase_diagram}.

{\bf Numerical results: } We test the dephasing mechanism and other predictions mentioned above in a long-range mixed field Ising model with Hamiltonian
\begin{equation}
H = - \sum \limits_{r,r'} \frac{J}{|r-r'|^\alpha }\sigma^z_r \sigma^z_{r'} - \sum\limits_{r} h_z \sigma^z _r -\sum\limits_{r} h_x \sigma^x _r,
\end{equation}
where $J$ is set to 1 as the energy unit, and the fields $h_z$ and $h_x$ are set to $0.5$ and $1.05$, respectively.

We implement the TDVP algorithm in operator space, which treats the operator as a matrix-product state and optimizes within the space of matrix-product representations \cite{Haegeman2011, Haegeman2016, Leviatan2017}. 
The ``super'' Hamiltonian $\mathcal{H}=H\otimes I - I\otimes H^*$ of the long-range interaction is explicitly constructed and fed into the state-based TDVP algorithm~\cite{Haegeman2016}. 
We expect that information far ahead of the wave front can be extracted with relatively low bond dimension, enabling us to simulate up to $200$ sites.


\begin{figure}
\subfigure[]{
  \label{fig:linear_cone}	
  \includegraphics[width=0.46\columnwidth]{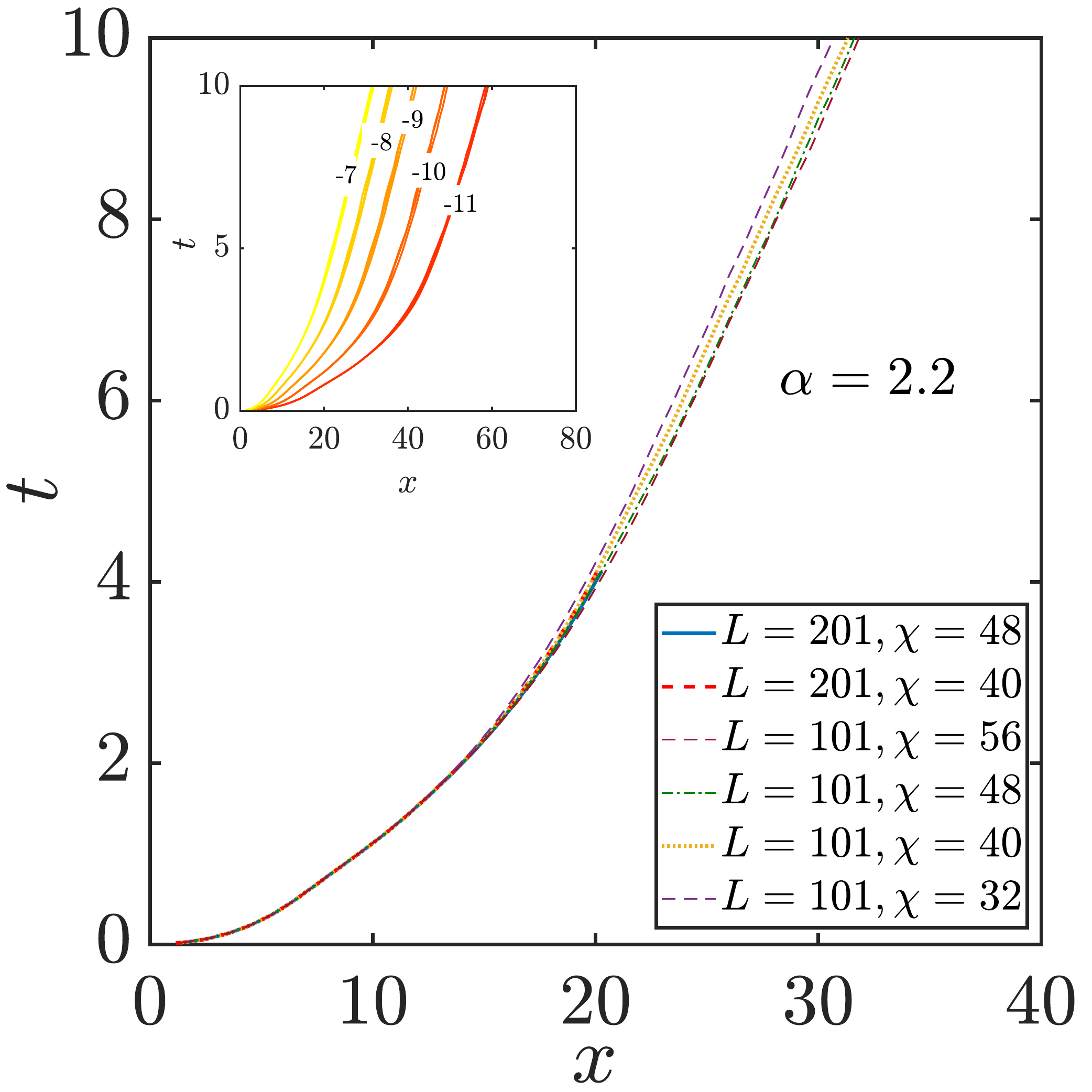}
}
\subfigure[]{
  \label{fig:power_cone}	
  \includegraphics[width=0.46\columnwidth]{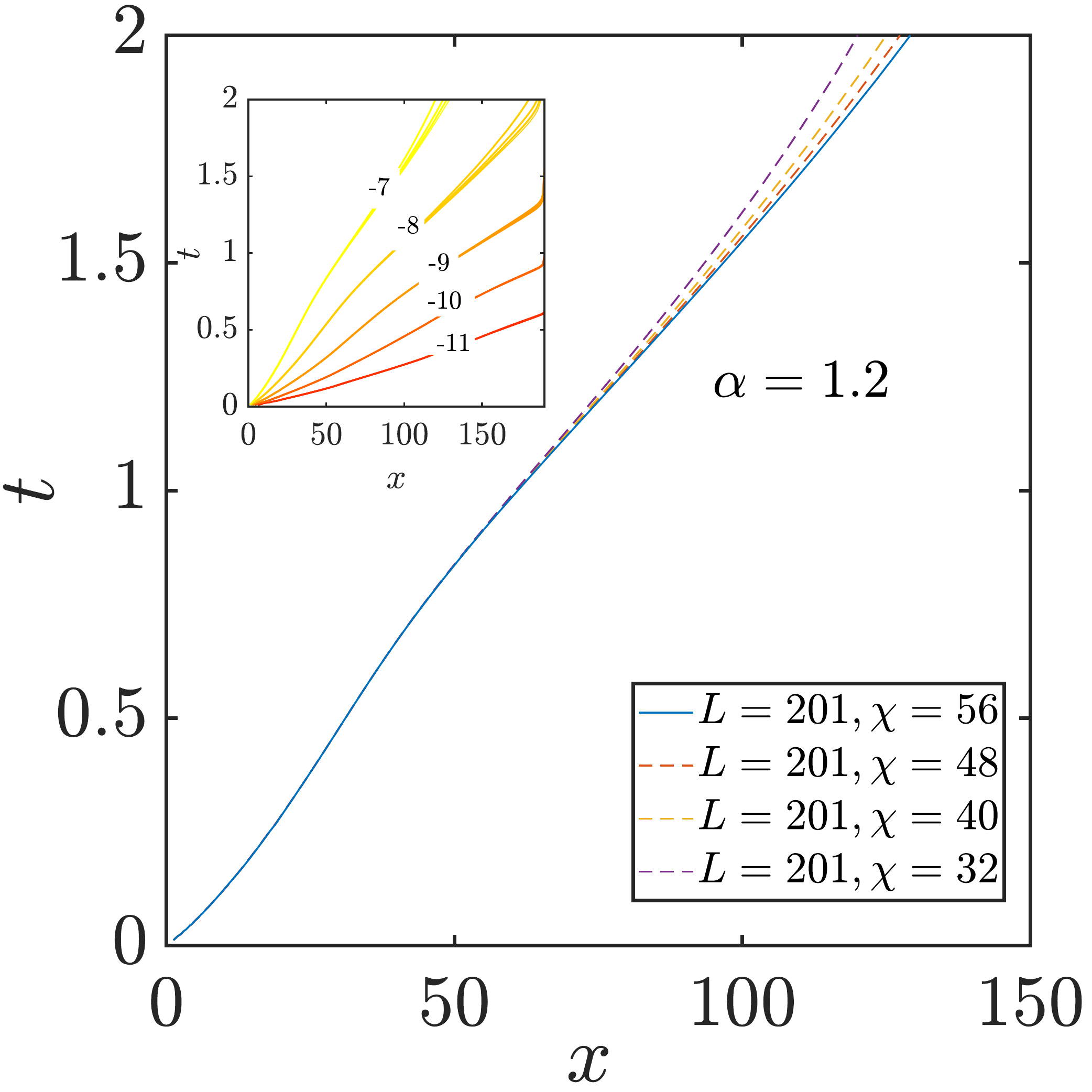}
}
\caption{The light cone of the long-range mixed-field Ising model for (a) $\alpha = 2.2$ and (b) $\alpha = 1.2$. Contours of $C(x, t)$ at threshold $\epsilon = e^{-7}$ are the main figures and other thresholds in the insets. Various system sizes and bond dimensions confirm convergence. 
}
\label{fig:num_lightcone}
\end{figure}

In Fig.~\ref{fig:num_lightcone}, we present the contour plots of $C(x,t)$ for $\alpha = 2.2$ and $\alpha = 1.2$, which demonstrate the linear and power-law light cones respectively. The insets show the contours for different values of the threshold, $\epsilon$. Eq.~\eqref{eq:rho_monep} predicts that the contours will follow the relations $(x - v_Bt) / \sqrt{t} \sim \text{constant}$ and $x \sim t^{\frac{1}{\zeta}}$ for the linear and power-law light cones respectively. The former gives convex curves that become parallel asymptotically, while the latter gives concave curves that disperse. These features are reflected in Fig.~\ref{fig:linear_cone} and Fig.~\ref{fig:power_cone}.


\begin{figure}
\subfigure[]{
  \label{fig:pt_init}	
  \includegraphics[width=0.46\columnwidth]{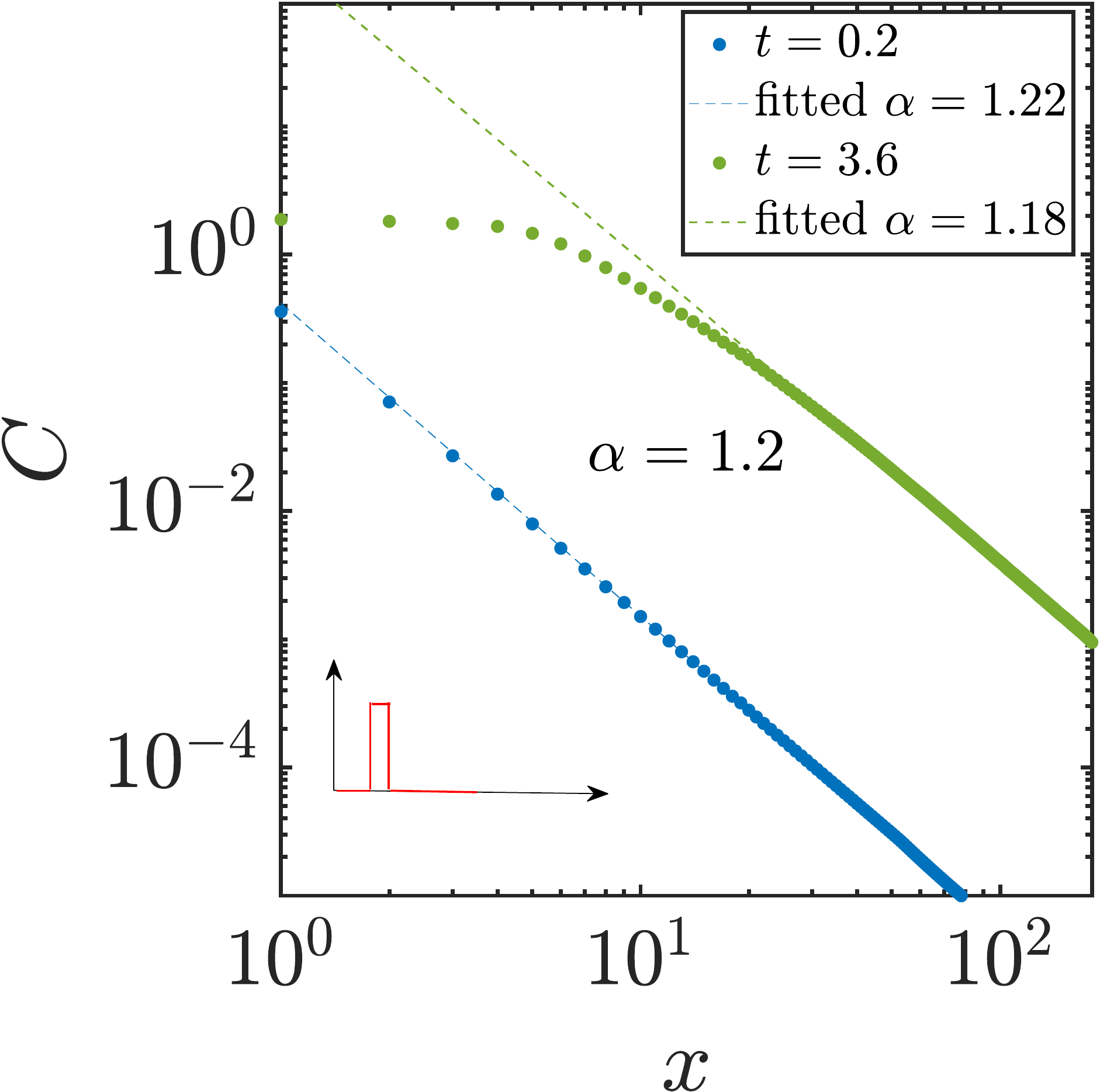}
}
\subfigure[]{
  \label{fig:dw_init}	
  \includegraphics[width=0.46\columnwidth]{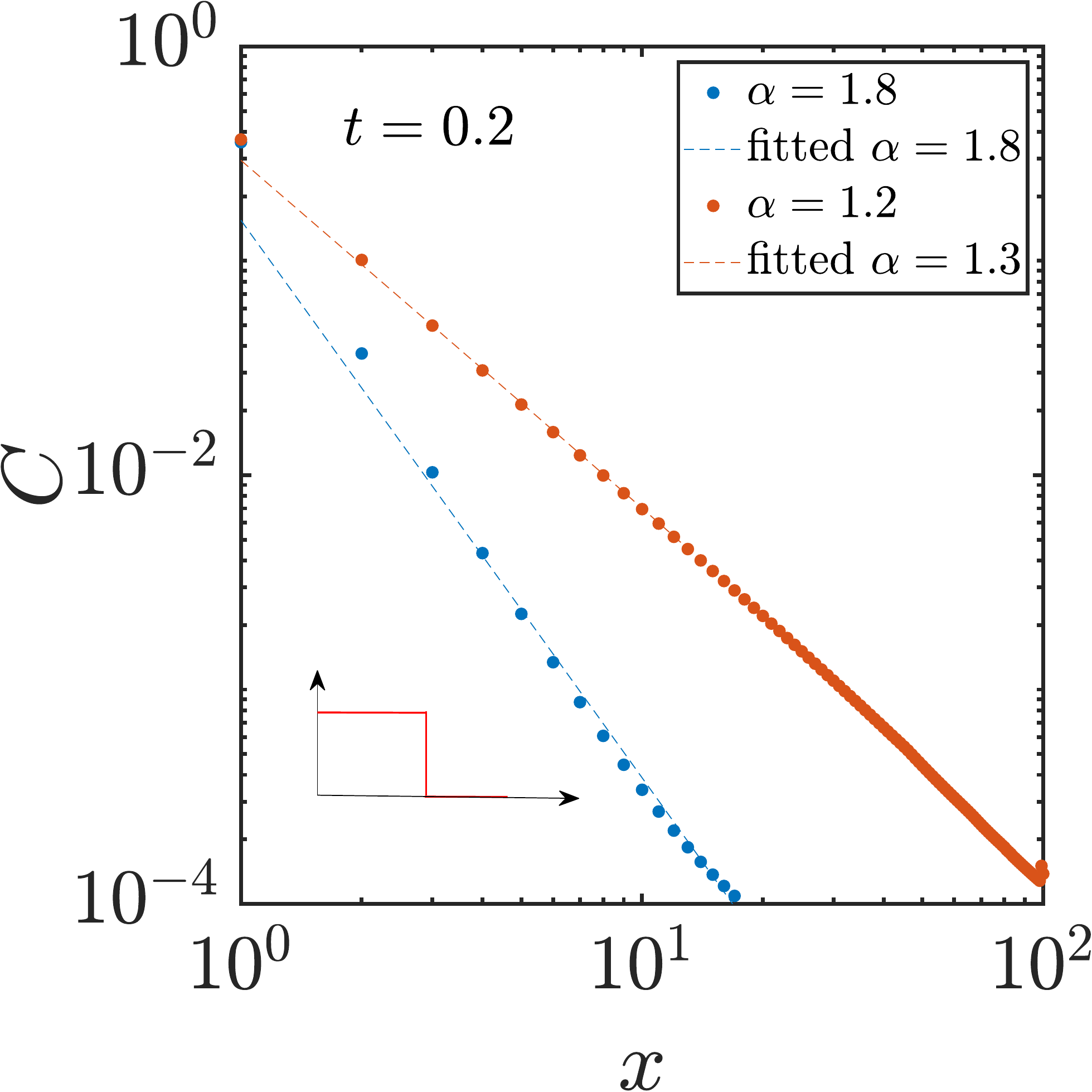}
}
\caption{Tail of the front for (a) a point and (b) domain wall initial conditions. (a) at $\alpha = 1.2$, the decay fits $x^{-2\alpha}$ at long times. (b) the short time decay fits $C=a\left (x^{1-2\alpha_{\rm fitted}}-(x+x_0)^{1-2\alpha_{\rm fitted}} \right )$ where $x_0$ is the domain wall length. $\alpha_{\rm fitted} \approx \alpha$, confirming the L\'evy flight prediction. 
}
\label{fig:tail_scaling}
\end{figure}

A precise verification of the phase boundary is computationally challenging. We instead measure the spatial dependence of the power-law tail to verify the proposed dephasing scheme. Fig.~\ref{fig:pt_init} shows the tail of the front for a point initial condition with $\alpha=1.2$. The decay exponent remains close to $2\alpha$ even at {\it late} times, consistent with the mean field argument \cite{SM}.
In contrast, a domain wall initial condition with $h  = 1$ for $x<0$ will generate a tail that scales as ${x^{-2\alpha - 1 }}$ at {\it early} times. In Fig.~\ref{fig:dw_init}, we fit the decay while taking into account the finite size of the domain and show that the fitting parameter $\alpha_{\rm fitted}$ is fairly close to $\alpha$.


{\bf Discussion and conclusion: }
We studied information propagation in chaotic long-range interacting systems via an analysis of the light cone structure of the squared commutator. Invoking a dephasing mechanism, we proposed a general phase diagram for such chaotic systems that generalizes the one proposed in Ref.~\cite{chen_quantum_2018} that exhibits logarithmic, power-law and linear light cone regimes. In particular, we analytically compute and numerically confirm the emergence of a linear light cone when the power-law exponent of the interaction strength $\alpha \ge 1.5$. { The powerful TDVP-MPO algorithm allows us to simulate systems with 200 sites, so that pertinent results at late times can be explicitly verified.}

A further simplification of the model yields a simple L\'evy flight picture (\monep) that describes the operator spreading in generic long-range interacting systems. It is remarkable that we can determine all the phase transition points at where the moments of L\'evy flight diverge, as well as the OTOC scaling close to the light cone. Both \mone and the associated arguments are also generalizable to systems with a large number of on-site degrees of freedom, which we leave to future work. 

Recently, Ref.~\cite{chen_finite_2019} proved a general Lieb-Robinson-type bound with a linear light cone for $\alpha > 3$ in 1d.
We here have a smaller threshold at $\alpha = 1.5$. This is in accordance with folklore that chaos usually prevents a optimal rate of propagation.
Thus, we anticipate that the critical $\alpha$ for the systems we consider will generally be smaller than those of theoretical bounds.


{\bf Acknowledgements:}
We acknowledge insightful discussions with Minh Tran and especially Sarang Gopalakrishnan for pointing out the relevance to the L\'evy flight at very early stage of the project.
We also thank the accommodation and interactive environment of the KITP program ``The Dynamics of Quantum Information" and the Aspen winter conference ``Many-Body Quantum Chaos".
XC and TZ are supported by postdoctoral fellowships from the Gordon and Betty Moore Foundation, under the EPiQS initiative, Grant GBMF4304, at the Kavli Institute for Theoretical Physics.
XC acknowledges support from DARPA DRINQS program. This research is supported in part by the National Science Foundation under Grant No. NSF PHY-1748958.
We acknowledge support from the Center for Scientific Computing from the CNSI, MRL: an NSF MRSEC (DMR-1720256) and NSF CNS-1725797, and University of Maryland supercomputing resources.
S. X and B. S acknowledge support from the U.S. Department of Energy, Office of Science, Advanced Scientific Computing Research Quantum Algorithms Teams program as part of the QOALAS collaboration.
AYG is supported by the NSF Graduate Research Fellowship Program under Grant No. DGE-1840340.
AYG also acknowledges partial support by the DoE ASCR Quantum Testbed Pathfinder program (award No. DE-SC0019040), DoE BES QIS program (award No. DE-SC0019449), NSF PFCQC program, AFOSR, ARO MURI, ARL CDQI, and NSF PFC at JQI.

\appendix

\section{Discrete L\'evy Flight and the Generalized Central Limit Theorem}
\label{app:GCL}

In this section, we review elementary results about L\'evy flight~\cite{calvo_generalized_2010,janson_stable_2011,chechkin_introduction_2008}. 

The L\'evy flight is a long range random walk. Its displacement at each step is independently drawn from a distribution $f(x)$ that has an asymptotic power law tail:
\begin{equation}
f(x) \rightarrow \frac{c_{\pm} }{x^{1+\alpha}}  \text{ for } x \rightarrow \pm \infty.
\end{equation}
If the second moment of the distribution exists ($\alpha>2$), then according to the central limit theorem, the total displacement will converge to a standard normal distribution with mean $v_B t$, where $v_B$ is the first moment.

The generalized central limit covers the cases when the second moment does not exist. Specifically, let $\{x_1, x_2, \cdots, x_t\}$ to be the independent displacements of the L\'evy flight, then\footnote{When $\alpha = 2$, $Y$ should be defined as $ \frac{x_1 + x_2 + \cdots + x_t }{(t\ln t)^{\frac{1}{\alpha}}} $, which converges to a normal random variable.} the rescaled displacement $Y = \frac{x_1 + x_2 + \cdots + x_t }{t^{\frac{1}{\alpha}}} $ converges to a random variable with distribution $L_{\alpha, \beta} ( y;  \frac{v_B t }{t^{\frac{1}{\alpha}}}  , \sigma_0 ) $.

$L_{\alpha, \beta}( x; \mu, \sigma) $ is the L\'evy stable distribution defined through its characteristic function
\begin{equation}
\Psi( k ) = \exp\left[ i \mu k  - \sigma^\alpha | k |^{ \alpha} \left( 1 - i \beta \text{sgn}(k) \omega( k , \alpha ) \right)  \right].
\end{equation}
Here $\mu$ is the first moment (which equals $v_B$ in our case), $\sigma$ is the scale parameter (a generalization of variance),  $ - 1 \le \beta = \frac{c_+ - c_-}{c_+ + c_-} \le 1$ is the skewness parameter defined by the asymptotic decays of the distribution, and 
\begin{equation}
\label{eq:omega_rule}
\omega( k, \alpha ) = \left\lbrace
  \begin{aligned}
    & \tan \left( \frac{\pi \alpha}{2} \right)  & \quad \alpha \ne 1, \\
    & -\frac{2}{\pi} \ln | k |  & \quad \alpha = 1. \\
  \end{aligned} \right.
\end{equation}

Through change of variable, the total displacement $\sum_{i=1}^t x_i$ scales as $ \frac{1}{\sigma_0 t^{\frac{1}{\alpha}}} L_{ \alpha \beta }( \frac{x - v_B t }{\sigma_0 t^{\frac{1}{\alpha}}} )$ (when $\alpha < 1$, we can set $v_B = 0$). The L\'evy stable distribution decays as $x^{-(1+ \alpha) }$, i.e. the same scaling as those long jumps.

The L\'evy distribution we use in the text has power law exponent $2\alpha - 2$ and skewness parameter $1$.


\section{The tail scaling analysis for \mone and \monep}
\label{app:tail}

In this section, we compare the tail distributions (of the front) in \mone and \monep. 

The tail distribution of \monep $\rho(x, t)$ is determined by the right-most point. It performs a L\'evy flight that has the same tail distribution as the jump distribution $f_{\rm jump} (x )$. In other words $\rho(x, t)$ has the same tail as $f_{\rm jump} (x )$. Finally $C(x,t)$ is the probability for site $x$ to be occupied, hence should corresponds to the cumulant distribution of the $\rho(x,t)$
\begin{equation}
C(x,t) = \int_x^\infty \rho(x', t ) \,dx'.
\end{equation}
With the explicit expression of $\rho(x, t)$, we obtain the tail distribution of \monep in Tab.~\ref{tab:op_tail}. 
\begin{table}[h]
\centering
\begin{tabular}{ |c|c|c|c|c| }
 \hline
  $\alpha$ & \monep tail  & \mone tail \\ \hline
  $ 1 < \alpha \le \frac{3}{2} $ & $x^{- (2\alpha - 2)}$ & $x^{- 2\alpha}$\\ \hline
  $ \frac{3}{2} < \alpha < 2 $ & $x^{- (2\alpha - 2)}$ & $x^{- (2\alpha - 2)}$ \\ \hline
  $ \alpha = 2 $ &  Gaussian &  Gaussian\\ \hline
  $ 2 < \alpha $ &  Gaussian &  Gaussian\\ \hline
\end{tabular}
\caption{Tail scalings of \monep and \mone. For $1 < \alpha < 2$, the L\'evy flight has tail $\rho( x', t) \sim \frac{1}{x^{2 \alpha - 1}}$. So $C(x,t) = \int_{x'}^{\infty}\rho(x', t ) dx' $ has tail $\frac{1}{x^{2\alpha - 2}}$. \mone has identical data except that the tail for $1 < \alpha < 1.5$ scales as $\frac{1}{x^{2\alpha}}$ (see text).  }
\label{tab:op_tail}
\end{table}

The tails of \mone and \monep will have the same scaling when a domain of occupied sites exists.

We only expect their behaviors to differ for $0.5 < \alpha < 1.5$. Taking a point $x$ far away from the light cone, for $0.5 < \alpha < 1.5$, this means $x / x_{\rm LC}(t)  \gg  1$. One expect that the occupied sites in each instance of \mone are scattered outside the light cone rather than forming a contiguous domain. Hence $C(x, t)$ should be roughly the jump rate within the light cone to the site at $x$. On large scales, we use the mean field approximation to estimate
\begin{equation}
\begin{aligned}
C(x, t ) &\sim \sum_{|x'|< x_{\rm LC}(t)} C(x' , t ) \frac{1}{( x - x')^{ 2\alpha } }   \\
 &\sim  \frac{1}{( x - x_{\rm LC}( t))^{2\alpha} } \sum_{|x'|< x_{\rm LC}(t)} C(x' , t ).
\end{aligned}
\end{equation}
In this regime, the $\text{Log}^{\frac{1}{\eta}}$ and power-law light cone suggest that $C(x,t)$ is scale free. We thus use a power-law ansatz  $C(x, t ) \sim \left( \frac{x}{x_{\rm LC}} \right)^{-\alpha_{\rm tail} }$. It gives $ C(x,t) \sim  \frac{1}{x^{ 2\alpha } }$ after plugging in, which implies $\alpha_{\rm tail} = 2\alpha$. The scaling of $C(x, t) \sim \left( \frac{x}{x_{\rm LC}} \right)^{-2\alpha }$ is consistent with the previous numerical study of \mone\cite{chen_quantum_2018} for $0.5 < \alpha < 1.5$.


\section{Brownian Circuit and its numerical data}
\label{app:bc}

The Brownian circuit is a model that contains only noisy interactions~\cite{zhou_operator_2018,chen_quantum_2018,xu_locality_2018,lashkari_towards_2013}. Hence the evolution of $f(\vec{h})$ is a Markov process. In 1d, we have the following master equation ~\cite{zhou_operator_2018,chen_quantum_2018,xu_locality_2018} (also see the full derivation in the next section)
\begin{equation}
\label{eq:master-N1}
\begin{aligned}
\frac{\partial f( \vec{h}, t ) }{\partial t}  =& \sum_{j \ne i }  3 D_{ij} h_{j} f( \vec{h} - \vec{e}_i , t ) +  \sum_{ j \ne i } D_{ij} h_{j} f( \vec{h} + \vec{e}_i , t ) \\
 & - \left\{ \sum_{j\ne i } 3 D_{ij} h_{j} (1-h_i) +  D_{ij}   h_i  h_{j} \right\} f( \vec{h}, t )  .
\end{aligned}
\end{equation}
The first two terms describe the transition rates from a height configuration $\vec{h} \pm  \vec{e}_i$ to $\vec{h}$, where the component of $\vec{e}_i$ is $1$ at site $i$ and $0$ elsewhere. The coefficients $D_{ij} = \frac{1}{|i - j|^{2\alpha}}$ is proportional to the square of the quantum interaction strength -- dephasing mechanism is at work here. If we take the local Hilbert space to be $q$-dimensional, then the transition rate should be replaced by $4 (1 - \frac{1}{q^2}) D_{ij}$ and $\frac{4}{q^2}D_{ij}$. The transition of height decrease, i.e. the $f( \vec{h} + \vec{e}_i, t) $ term, has a coefficient suppressed by $\frac{1}{q^2}$. In the $q\to\infty$ limit it vanishes and we get \mone.

\subsection{The numerics of Brownian Circuit / \mone}
\label{app:tail}

\begin{figure}[h]
\centering
\subfigure[]{\label{fig:alpha_14_seed_collapse} \includegraphics[width=.75\columnwidth]{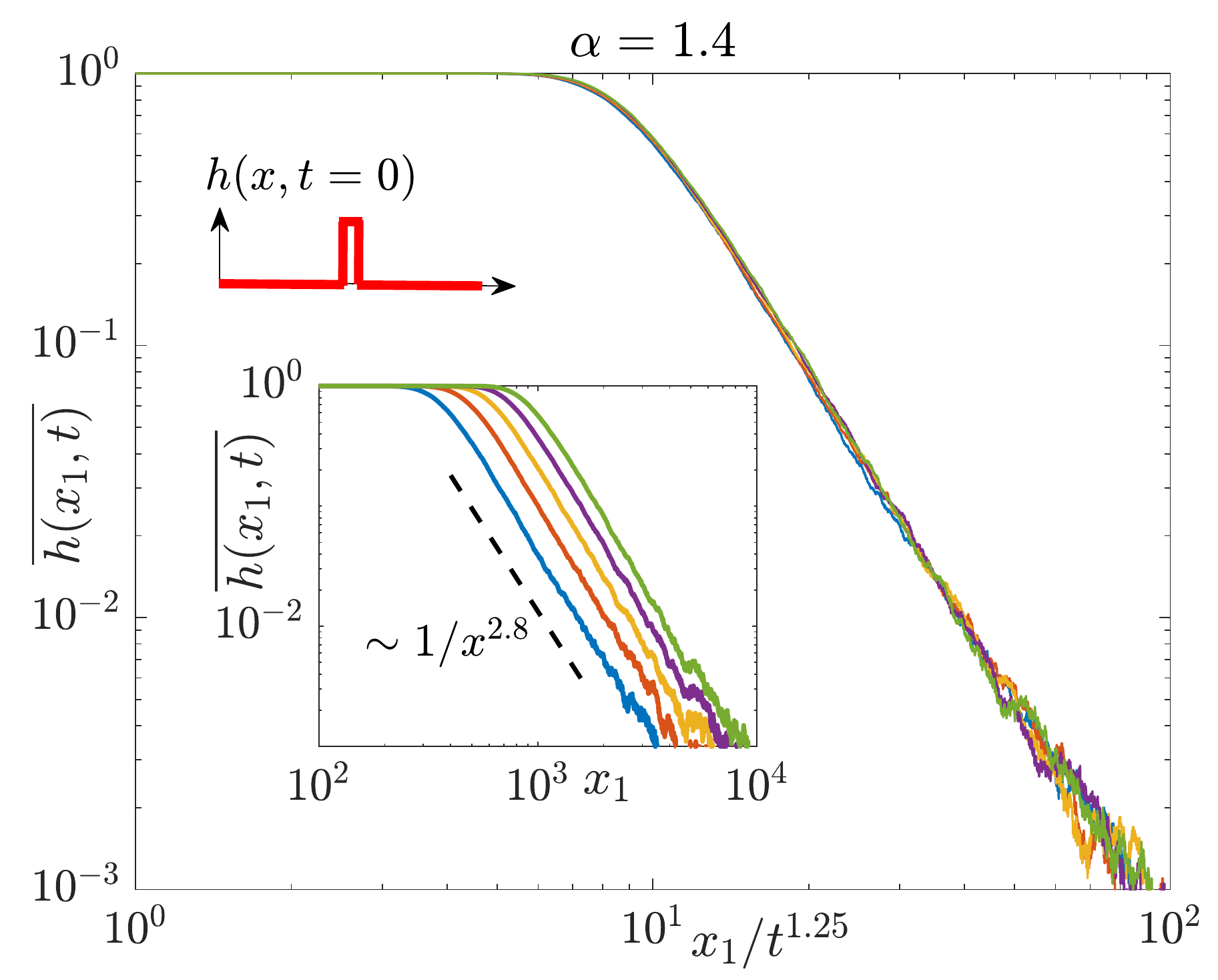}}
\\
\subfigure[]{\label{fig:alpha_12_seed_collapse} \includegraphics[width=.75\columnwidth]{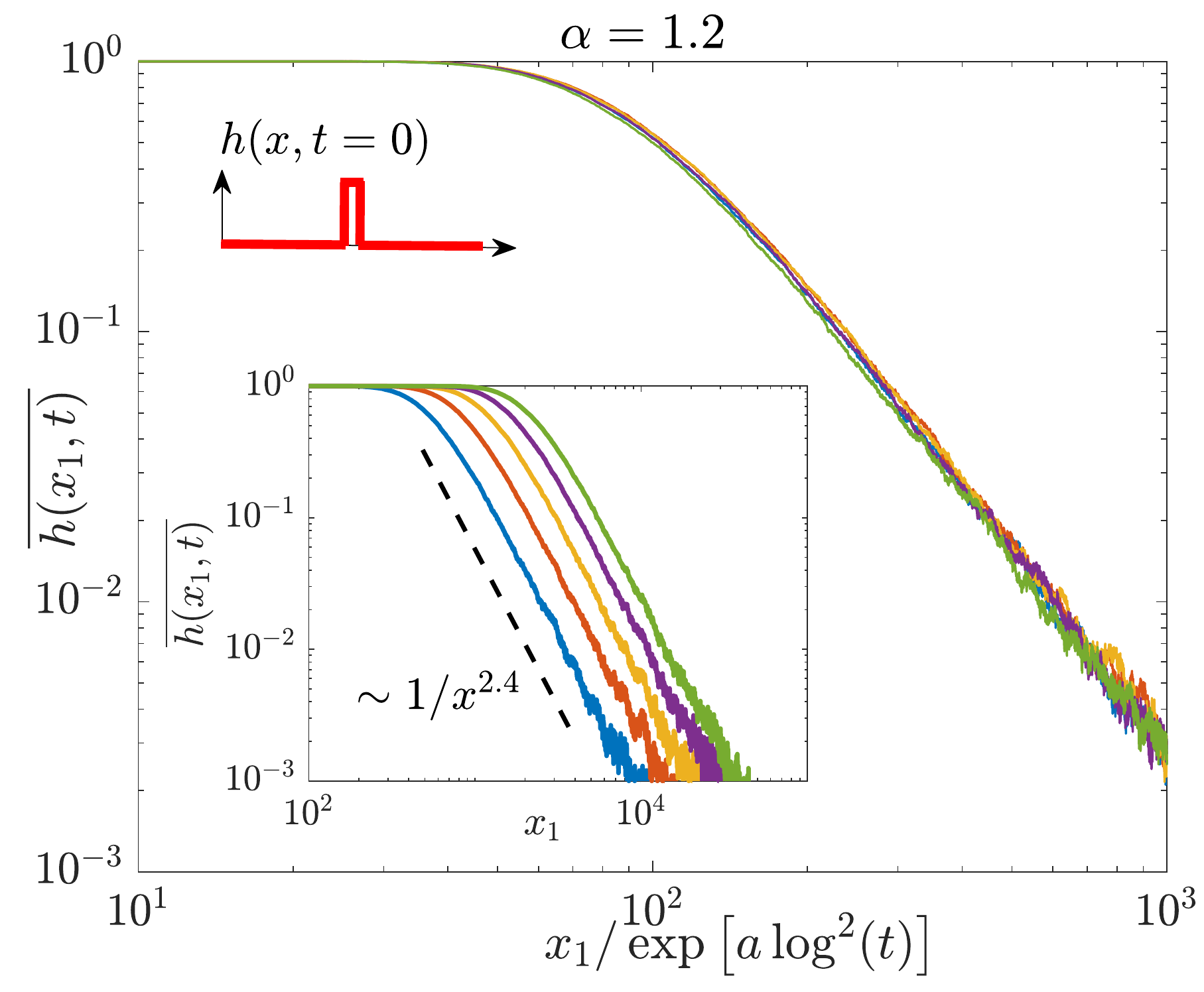}}
\caption{ The data collapse of $\overline{h(x,t)}$ for various $\alpha$ with $L=100,000$. The mean height  $\overline{h(x,t)}$ is obtained after taking average over 20,000 simulations. The initial condition is taken as  the Kronecker delta function $h(x,t)=\delta_{x,L/2} $ with $x_1\equiv x-L/2$. (a) When $\alpha=1.4$, we take the scaling argument to be $x_1/t^{1.25}$, consistent with the theoretical prediction in Eq.\eqref{eq:light_cone}.  (b) When $\alpha=1.2$, we choose the scaling argument to be $x_1/\exp\left[ a\log^2(t)\right]$, which is supposed to be working for $\alpha=1$. Here we take $a=0.42$ which is larger than $1/4\log(2)$.}
\label{fig:collapse_front}
\end{figure}

Refs.~\cite{hallatschek_acceleration_2014,chatterjee_multiple_2013} proved the asymptotic light cone structures of \mone. 
In one dimension, they read:
\begin{align}
 x_{\text{LC}} \sim
  \begin{cases}
    t       & \quad  1.5 \le \alpha \\
   t^{\frac{1}{2\alpha - 2}}  & \quad 1<\alpha<1.5 \\
   \exp\left[ \frac{1}{4 \log 2} \log^2 (t)\right]  & \quad \alpha=1 \\
   \exp(B_\alpha t^\frac{\log \frac{1}{\alpha} }{\log  2})  & \quad 0.5 \le \alpha<1
  \end{cases}.
  \label{eq:light_cone}
\end{align}
The power-law light cone regime between $1$ and $1.5$ is the same as that for \monep.

We numerically check the power-law light cone scalings. In Fig.~\ref{fig:alpha_14_seed_collapse}, the data collapse of the mean height $\overline{h(x,t)}$ with the scaling arguments $x/t^{1/(2\alpha-2)}$ is very successful for $\alpha = 1.4$. However, as emphasized by Ref.~\cite{hallatschek_acceleration_2014}, it converges very slowly to the power-law light cone when $\alpha \rightarrow 1$. In fact, when $\log t \le \frac{2d}{|\alpha - 1|}$, the light cone scaling will flow to the marginal case of $\alpha = 1$. As a result, we collapse the $\alpha = 1.2$ case with scaling argument  $x/\exp\left[ a\log^2(t) \right]$ in Fig.~\ref{fig:alpha_12_seed_collapse}.

\begin{figure}[h]
\centering
\includegraphics[width=.75\columnwidth]{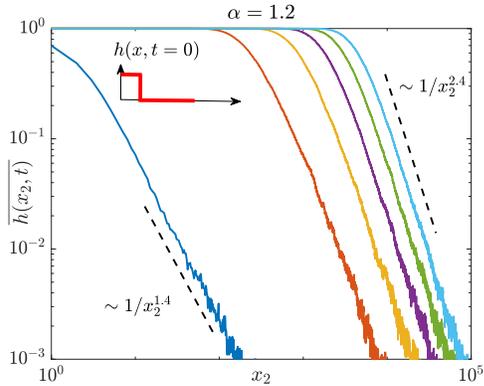}
\caption{ The height dynamics with the domain wall initial condition: $h(x<1000,t=0)=1$ and $h(x>1000,t=0)=0$. As time evolves, the exponent of the power law tail changes from $1.4$ to $2.4$. Here the mean height is obtained after taking average over 20,000 simulations and the total system size is $L=100,000$.}
\label{fig:alpha_12_domain_wall} 
\end{figure}

Finally, we numerically check the shape of the front. Starting from an initial condition which takes nonzero value only in the middle of system, we have $\overline{h(x)}\sim 1/x^{2\alpha}$ ahead of the light cone, as shown in the insets of Fig.~\ref{fig:alpha_14_seed_collapse} and Fig.~\ref{fig:alpha_12_seed_collapse}. Additionally starting from a domain wall initial condition, we observe the crossover from $1/x^{2\alpha-1}$ scaling to $1/x^{2\alpha}$ scaling (see Fig.~\ref{fig:alpha_12_domain_wall}). In the long time limit, we always have the $1/x^{2\alpha}$ scaling behavior.

\section{Master Equation of height in Brownian circuit}
\label{app:master_eq}

In this section, we give a detailed derivation of the master equation in Brownian circuit.

We allow the Hamiltonian to have general two-body interaction in a local $q$-dimensional Hilbert space
\begin{equation}
d G(t) = \sum_{i <j} A_{ij} dB(t)_{i,j,\mu_i, \mu_j} \sigma^{ \mu_i}_i \sigma^{\mu_j}_j,
\end{equation}
where we label each spin by lowercase Roman indices $i, j, k$. Here $\sigma^{ \mu _i}_i$  are set of Hermitian basis for $i$th spin, which are chosen to be
\begin{equation}
  \sigma^{\mu}
   = \left\lbrace
     \begin{aligned}
       & \I_q  & \quad \mu = 0 \\
       & \sqrt{ 2 q } T_a & \quad \mu = a > 0 \\
     \end{aligned} \right. .
 \end{equation}
$T_a$ are the standard $\text{SU}(q)$ generators. They are $q^2 - 1$ traceless Hermitian matrices normalized as
 \begin{equation}
 T_a T_b =  \frac{1}{2q}  \delta_{ab} \I_q  + \frac{1}{2} \sum_{c= 1}^{ q^2 - 1 } ( d_{ab}{}^c +  i f_{ab}{}^c  ) T_c
\end{equation}
so that
\begin{equation}
\tr( \sigma^{ \mu } \sigma^{\nu} ) = \tr( \I_q) \delta_{\mu\nu}  = q \delta_{\mu \nu }.
\end{equation}

For time-dependent noisy dynamics, we should expand the evolution to second order (\cf Lindblad equation) and apply the It\^{o} formula,
\begin{equation}
\begin{aligned}
  dO(t)  &=  [ i dG(t), O (t) ]  + \frac{1}{2} [ i dG(t), [ i dG(t), O (t) ] ] \\
  &= i [ dG(t), O(t) ] - \frac{1}{2} \{ dG(t) dG(t), O \}  + d G(t)  O d G (t) \\
  &= i [ dG(t), O(t) ] - r_0 O(t) dt + \sum_{i < j}  A_{ij}^2 q^2 \I_{ij} \tr_{ij} ( O ) dt,
\end{aligned}
\end{equation}
where in the last line we have used the following contraction identities:
\begin{equation}
\begin{aligned}
  dG(t) dG(t) &= r_0 \I dt, \quad r_0 =  \sum_{ i < j} A_{ij}^2 q^4, \\
  dG(t) O d G(t) &= \sum_{i < j } A_{ij}^2 q^2 \I_{ij} \tr_{ij} ( O ) dt.
\end{aligned}
\end{equation}
We are interested in the operator content of evolved operator $O(t)$. More precisely, let $B_{\mu}$ be the operator basis consisting of tensor products of $\sigma^{\nu}$ on each spin degree of freedom.
Letting $O(t) = \sum_{\mu } \alpha_{\mu}(t)  B_\mu$, we inspect the dynamical expansion coefficient
\begin{equation}
 \qquad \alpha_{\mu}(t) = \frac{1}{\tr( B_\mu B_\mu ) }\tr( B_{\mu} O(t) ).
\end{equation}
Its time evolution is given by:
\begin{equation}
\begin{aligned}
d \alpha_{\mu} (t) &= \frac{i}{\tr( B_\mu^2) }  \tr( B_{\mu} d O(t) ) \\
&= \frac{i}{\tr( B_\mu^2) }  \tr( [dG(t), O(t)] B_{\mu}  )  \\
&- r_0 \alpha_\mu (t) dt  +  q^4 \alpha_{\mu}(t) \sum_{B_{\mu}  \text{ is } \I \text{ on } i, j }  A_{ij}^2   dt\\
&=  \frac{i}{\tr( B_\mu^2  ) }  \tr( [dG(t), O(t)] B_{\mu} )  \\
&- \Big[ r_0 -  q^4  \Big( \sum_{B_{\mu}  \text{ is } \I \text{ on } i, j} A_{ij}^2 \Big) \Big] \alpha_\mu (t) dt.
\end{aligned}
\end{equation}
The first term is a noise term, whereas the second term is deterministic.

Define $f( B_\mu, t )$ to be the average probability at time $t$
\begin{equation}
f( B_{\mu} , t )  = \overline{|\alpha_{\mu}(t)|^2} = \overline{\alpha^2_\mu(t)}
\end{equation}
the evolution is given by
\begin{equation}
\begin{aligned}
  d f( B_\mu, t ) &= 2 \overline{\alpha_\mu(t) d \alpha_\mu(t) } + \overline{ d \alpha_\mu(t) d \alpha_\mu(t)  }.
\end{aligned}
\end{equation}
After doing the average, only the deterministic term will survive in the first differential and noisy term in the second differential. We have
\begin{equation}
\begin{aligned}
  d f( B_\mu, t ) &= - 2(\cdots ) \alpha^2_\mu (t) dt -  \frac{1}{\tr^2( B_\mu^2  )} \overline{\tr^2( [dG(t), O(t)] B_{\mu} )   }\\
  &= - 2(\cdots ) \alpha^2_\mu (t) dt -  \frac{1}{\tr^2( B_\mu^2  )} \overline{\tr^2( [B_\mu, dG(t)] O(t)  )   },
\end{aligned}
\end{equation}
where the dots represent $r_0 -   q^4  \Big( \sum_{B_{\mu}  \text{ is } \I \text{ on } i, j} A_{ij}^2 \Big)$. In a stochastic equation, this term can also be fixed by probability conservation, so we will not keep track of it. We can further reduce second term to other average probabilities
\begin{equation}
  \label{eq:df_general}
\begin{aligned}
  d f( B_\mu, t ) &= - ( \cdots )  f( B_\mu , t ) dt \\
  &-  \sum_{B_\nu}  \sum_{i<j} A^2_{ij}   \sum_{\mu_i, \mu_j}\frac{1}{\tr^2( B_\mu^2  )}  \\
  &\tr^2( [ B_\mu, \sigma_i^{\mu_i } \otimes \sigma_j^{\mu_j}] B_\nu )  f( B_\nu, t ) dt.
\end{aligned}
\end{equation}

At this point, the derivation is completely general about the spatial structure and the interaction types between those $q$-spins.

Now we specify the spatial structure and height variable. We use upper case roman index $I,J,K$ to label spatial sites. Each spatial site $I$ host $N$ spins. We define height variable on each site, and the joint height probability function $f( \vec{h}, t )$, where the vector $\vec{h}$ hosts height on each site. We assume equal partition on each local basis $\sigma^{\mu}$, then for any basis $B_{\mu}$ having height vector $\vec{h}$
\begin{equation}
  f( \vec{h}, t ) = f( B_\mu,  t )  C_{\vec{h} } \quad C_{\vec{h}} = \prod_I { N \choose h_I } (q^2 - 1)^{h_I }.
\end{equation}
We find that the 2-body interaction terms can only change the height by $\pm 1$, so can further restrict $B_{\nu}$ to $B_{\mu}^{+}$ and $B_{\nu}^{-}$. Thus we can multiply $C_{\vec{h}}$ on both sides of Eq.~\eqref{eq:df_general}
\begin{equation}
\begin{aligned}
  &d f( \vec{h}, t ) = -( \cdots ) f( \vec{h}, t ) dt  \\
  & - \frac{c_{\vec{h}}}{c_{\vec{h}- \vec{e}_I}}  \sum_{i < j} A_{ij}^2 \sum_{ B_\mu^{-} } \Delta_{ij}( B_{\mu}^{-} )  f( \vec{h} - \vec{e}_I , t ) dt
\\
  & - \frac{c_{\vec{h}}}{c_{\vec{h}+ \vec{e}_I}} \sum_{i < j}A_{ij}^2 \sum_{ B_\mu^{+} } \Delta_{ij}( B_{\mu}^{+} )  f( \vec{h} + \vec{e}_I , t ) dt
\end{aligned}
\end{equation}
where
\begin{equation}
  \begin{aligned}
    \Delta_{ij}( B_\mu^{-} ) &= \frac{1}{\tr^2( B_\mu^2  )} \sum_{\mu_i, \mu_j} \tr^2( [B_\mu, \sigma_i^{\mu_i } \otimes \sigma_j^{\mu_j} ]B_\mu^{-} ) \\
    \Delta_{ij}( B_\mu^{+} ) &= \frac{1}{\tr^2( B_\mu^2  )} \sum_{\mu_i, \mu_j} \tr^2( [B_\mu, \sigma_i^{\mu_i } \otimes \sigma_j^{\mu_j} ]B_\mu^{+} ) \\
    \frac{c_{\vec{h}}}{c_{\vec{h}- \vec{e}_I}} &= (q^2 - 1 ) \frac{N - h_I + 1}{h_I} \\
 \frac{c_{\vec{h}}}{c_{\vec{h}+\vec{e}_I}} &= \frac{1}{q^2 - 1} \frac{h_I + 1}{N - h_I}  .
\end{aligned}
\end{equation}
Notice that in the actual process, the transition from state $\vec{h} - \vec{e}_I$ to $\vec{h}_I$ induces a height {\it increase} rather than {\it decrease}. Our notation here refers to height decrease from basis $B_\mu$ to $B_{\mu}^{-}$

\begin{figure}[h]
\centering
\includegraphics[width=0.75\columnwidth]{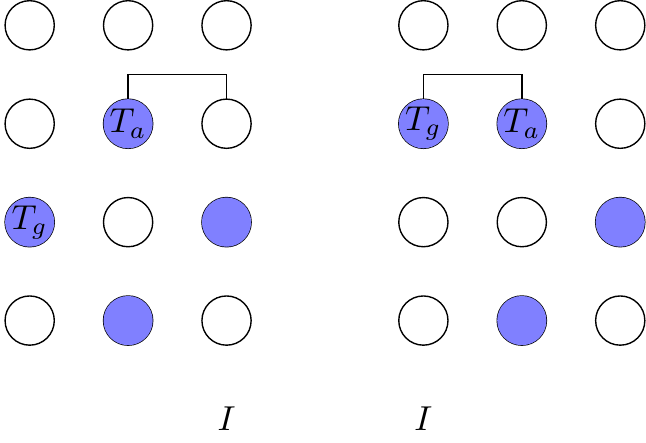}
\caption{Mechanism to change the height by $\pm 1$. Figure shows configuration for $B_{\mu}$, after the application of the interaction term (lines in the figure), it becomes $B_{\mu}^{+}$ (left) and $B_{\mu}^{-}$ (right). Left: Increase the height by $1$. One leg of the interaction must set foot on identity on site $I$: $( N- h_I) h_J$ choices to draw the lines. Right: Decrease the height by $1$. Both legs of the interaction must touch the non-identity: $ h_I h_J$ choices to draw the lines.}
\label{fig:hpm1}
\end{figure}

We now calculate the terms that change the height by $\pm 1$. First consider height increase. Then one leg of the interaction must be inside the basis and one outside, see left of Fig.~\ref{fig:hpm1}. We focus on one such interaction term, thus restricting to fixed spin $i$ and $j$
\begin{equation}
\begin{aligned}
 &\sum_{B_{\mu}^{+} } \Delta_{ij}( B_\mu^{+} ) =  \frac{1}{\tr^2( \I_q \otimes \I_q )} \sum_{b c h g}  \\
 &\tr^2( [ \sqrt{ 2q } T_a \otimes \I_q, \sqrt{ 2q }T_b \otimes\sqrt{ 2q } T_h] \sqrt{ 2q }T_c \otimes\sqrt{ 2q } T_g )  .
\end{aligned}
\end{equation}
Here we take a particular choice of $B_{\mu} = \sqrt{ 2q } T_a \otimes \I_q$ and sum over all possible choices of interactions $B_{\mu}^+ = \sqrt{ 2q }T_c \otimes\sqrt{ 2q } T_g$. Clearly, this can be reduced to one site case
\begin{equation}
\begin{aligned}
 &\sum_{B_{\mu}^{-} } \Delta_{ij}^{+}( B_\mu^{-} ) \\
 &=  \frac{1}{\tr^2( \I_q )} \sum_{b c}  \tr^2( [ \sqrt{ 2q }T_a , \sqrt{ 2q } T_b ]  \sqrt{ 2q } T_c  ) \sum_{hg} \delta^2_{hg}  \\
 &= \frac{q^2-1 }{\tr^2{(T_a^2) } } 2q \sum_{bc } \tr^2( [T_a , T_b ]  T_c ) \\
 &=  - 2q (q^2-1) \sum_{bc} f_{ab}{}^c f_{ab}{}^c = - 2(q^2 - 1 )q ^2,
\end{aligned}
\end{equation}
where we have used the $\text{SU}(N)$ identity
\begin{equation}
\begin{aligned}
  \sum_{bc} f_{ab}{}^c f_{ab}{}^c &= \sum_{bc} f_{bc}{}^a f_{bc}{}^a  = q \delta^{aa} = q \\
 & \text{ no summation on } a.
\end{aligned}
\end{equation}
There are $( N- h_I) h_J$ choices to create this type of interactions between site $I$ and $J$, if we assume $A_{ij} = J_{IJ}$ for all $i \in I$ and $j \in J$, then each choice contributes equally. The height increasing term becomes
\begin{equation}
  \begin{aligned}
    &\text{height increase} =
    - \frac{1}{q^2 - 1} \frac{h_I + 1}{N - h_I}( -2) ( q^2 - 1 ) q^2 \\
    &\qquad \sum_{i < j\text{ for } B_{\mu}^{-} }A_{ij}^2  f( \vec{h} + \vec{e}_I , t ) dt \\
    &=  2q^2 \frac{h_I + 1}{N - h_I} J_{IJ}^2 (N-h_I) h_J  f( \vec{h} + \vec{e}_I , t ) dt\\
    &= 2q^2 ( h_I +1 ) h_J J_{IJ}^2f( \vec{h} + \vec{e}_I , t ) dt.
\end{aligned}
\end{equation}

For height decrease, both legs of the interaction must touch the non-identities in $B_{\mu}$, see right of Fig.~\ref{fig:hpm1}. Again we reduce to two sites
\begin{equation}
\begin{aligned}
 &\sum_{B_{\mu}^{-} } \Delta_{ij}( B_\mu^{-} ) =  \frac{1}{\tr^2( \I_q \otimes \I_q )} \sum_{b h} \sum_{B_{\mu}^{-} } \\
&\tr^2( [ \sqrt{ 2q } T_g \otimes \sqrt{ 2q }T_a, \sqrt{ 2q }T_h \otimes\sqrt{ 2q } T_b]  B_{\mu}^{-})  .
\end{aligned}
\end{equation}
In the figure, we restrict site $I$ to host $T_g$ in $B_{\mu}$ and $T_h$ in the interaction term. In order for the height to decrease at site $I$, we must have $g = h $. Hence
\begin{equation}
\begin{aligned}
\sum_{B_{\mu}^{-} } \Delta_{ij}( B_\mu^{-} ) &= \frac{1}{\tr^2 ( \I_q ) }  \sum_{bc} \tr^2 ( [\sqrt{ 2q }T_a, \sqrt{ 2q }T_b]\sqrt{ 2q } T_c )   \\
&= - 2q^2  .
\end{aligned}
\end{equation}
Again, we assume that all interactions contribute to these two sites contributes equally. Then there are $h_I h_J$ choices. The height decreasing term becomes
\begin{equation}
\begin{aligned}
&\text{height decrease} = \\
&- (q^2 - 1) \frac{N - h_I + 1}{h_I} ( - 2q^2 ) h_I h_J J_{IJ}^2 f( \vec{h} - \vec{e}_I , t ) dt \\
&= 2q^2 ( q^2 - 1 ) ( N - h_I + 1) h_J J_{IJ}^2 f( \vec{h} - \vec{e}_I , t ) dt .
\end{aligned}
\end{equation}

Therefore overall we have
\begin{equation}
\begin{aligned}
  &d f( \vec{h}, t ) = -[2q^2 (q^2 - 1) \sum_{J} J_{IJ}^2 ( N - h_I ) h_J \\
  &\qquad \qquad + 2q^2 \sum_{J} J_{IJ}^2 h_I h_J] f( \vec{h}, t ) dt  \\
  &  +2q^2 (q^2 - 1) \sum_{J} J_{IJ}^2 ( N - h_I + 1 ) h_J  f( \vec{h} - \vec{e}_I , t ) dt  \\
  &  +2q^2 \sum_{J} J_{IJ}^2 (h_I +1 ) h_J  f( \vec{h} + \vec{e}_I , t ) dt.
\end{aligned}
\end{equation}

In the model we considered, we take $J_{IJ} = \sqrt{ \frac{2}{q^4} } \frac{1}{| I - J|^\alpha}$. This normalization gives
\begin{equation}
\begin{aligned}
  & d f( \vec{h}, t ) = -[4 (1 - \frac{1}{q^2} )  \sum_{J} \frac{1}{|I - J|^{2\alpha} } ( N - h_I ) h_J  \\
  &+ \frac{4}{q^2} \sum_{J} \frac{1}{|I - J|^{2\alpha} } h_I h_J] f( \vec{h}, t ) dt  \\
  &  + 4 (1 - \frac{1}{q^2} )  \sum_{J} \frac{1}{|I - J|^{2\alpha} } ( N - h_I + 1 ) h_J  f( \vec{h} - \vec{e}_I , t ) dt  \\
  &  + \frac{4}{q^2} \sum_{J} \frac{1}{|I - J|^{2\alpha} } (h_I +1 ) h_J  f( \vec{h} + \vec{e}_I , t ) dt.
\end{aligned}
\end{equation}



\section{TDVP Method for Numerical Simulation}
\label{app:num}

\def\be{\begin{equation}}
\def\ee{\end{equation}}
\def\bea{\begin{equation}\begin{aligned}}
\def\eea{\end{aligned}\end{equation}}
\def\PRB{Phys. Rev. B}
\def\PRA{Phys. Rev. A}

\def\avg#1{\langle#1\rangle}
\def\Re{\rm{Re}}
\def\Im{\rm{Im}}
\def\PRL{Phys. Rev. Lett.}
\def\nn{\nonumber}
\def\pp{\parallel}

\def\w{\mathbf{w}}
\def\e{\mathbf{e}}
\def \erf {\text{erf}}
\def \tr {\text{tr}}

In this section, we give the detailed construction of the numerical method (TDVP-MPO). The basic idea is to treat the operator as a quantum state in matrix product operator forms and evolve it using the time dependent variational approach. 

Consider the following generic long-range Hamiltonian,
\begin{equation}\begin{aligned}
H=\sum \limits_{r,r', \alpha, \beta} V^{\alpha\beta}(r-r') O_r^\alpha O_{r'}^\beta + \sum\limits_{r} h^\alpha _r O^\alpha _r.
\label{eq:long_range_ham}
\end{aligned}\end{equation}
The corresponding super-Hamiltonian that describes the operator dynamics is $\mathcal{H} = H\otimes I - I \otimes H^*$. We write the super-Hamiltonian in a matrix product form,
\begin{equation}\begin{aligned}
\hat{H} = \hat{V}_l \hat{M}_1 \hat{M}_2 \hat{M}_3 ... \hat{M}_L \hat{V}_r,
\end{aligned}\end{equation}
where $V_{l/r}$ is the boundary vector of operators (each element of the vector is an operator) and $M$s are the matrices of operators defined on each site. The boundary vector, and operator matrices can be  constructed explicitly for the long-range super Hamiltonian given in Eq.~\ref{eq:long_range_ham}. The on-site term have a simple bond dimension 2 MPO representation:
\begin{equation}\begin{aligned}
&\hat{V}_l = (0, \hat{I}\otimes \hat{I}), \ \ \hat{V}_r = (\hat{I}\otimes \hat{I}, 0)  \\
&\hat{M} =
\begin{pmatrix}
\hat{I}\otimes \hat{I} & 0 \\
h^\alpha \left ( O^\alpha_r \otimes \hat{I} - \hat{I}\otimes O_r^{\alpha*} \right )  & \hat{I}\otimes \hat{I} \\
\end{pmatrix}.
\end{aligned}\end{equation}
On the other hand, the long-range term between single pair of operators $\sum \limits_{r,r'} V^{\alpha\beta}(r-r') O_r^\alpha \otimes I_r  O_{r'}^\beta \otimes I_{r'}$, appearing in the super Hamiltonian $\mathcal{H}$ has the following MPO form with $L+1$ dimensional boundary vectors and $L+1 \times  L+1$ dimensional operator matrices,
\begin{equation}\begin{aligned}
&\hat{V}_l ^{L+1}=\hat{I}\otimes \hat{I},  \ \ \hat{M}_i^{1,1}= \hat{I}\otimes \hat{I} \\
 &\hat{M}_i^{a+1 , 1}= h(a)O_i^\beta \otimes \hat{I} , \ \ \hat{M}_i^{ 1+a ,  2+a} = \hat{I} \otimes \hat{I},  \ \  \\&(a=1, 2, ... , L-1) \\
&\hat{M}_i^{L+1, 2} = \hat{O}_i^\alpha \otimes \hat{I},  \ \ \hat{M}_i^{L+1, L+1} = \hat{I}\otimes \hat{I},  \ \ \hat{V}_r^1 =\hat{I} \otimes \hat{I},
\end{aligned}\end{equation}
where the other entries are zero.

With all the pieces, the MPO for the whole Hamiltonian can be assembled together in a blocked form,
\begin{equation}\begin{aligned}
&V_l =( V_{l,1}, V_{l,2}, V_{l,3}, ... ),  V_r = ( V_{r, 1}, V_{r, 2}, V_{r,3}, ...) \\
&M = \begin{pmatrix}
M_1 &  0 & 0  & ...\\
0    &  M_2 & 0 \\
0  & 0  & M_3 \\
... & & & ...
\end{pmatrix},
\end{aligned}\end{equation}
which is ready to be used as the input in the time-dependent variational principle (TDVP) algorithm. Compression of the MPO, for example, via Schmidt decomposition, maybe required to reduce the memory usage.



\bibliographystyle{apsrev4-1}
\bibliography{op_levy_flight_paper}

\end{document}